\documentclass[pra,twocolumn,a4paper,showpacs,superscriptaddress,prl]{revtex4}
\usepackage{amsmath}
\usepackage{amsfonts}
\usepackage{graphicx}
\begin{document}
\title{Characterizing  multiparticle entanglement in symmetric $N$-qubit states \\  
via negativity of covariance  matrices} 
\author{A. R. Usha Devi}
\email{arutth@rediffmail.com}
\author{ R. Prabhu}
\affiliation{Department of Physics, Bangalore University, 
Bangalore-560 056, India}
\author{A. K. Rajagopal}
\affiliation{Department of Computer Science and Center for Quantum Studies, George Mason 
University, Fairfax, VA 22030, USA and
Inspire Institute Inc., McLean, VA 22101, USA.} 

\date{\today}

\begin{abstract}
We show that higher order  inter-group covariances  involving even number 
of qubits  are necessarily positive semidefinite for 
$N$ qubit separable states, which are completely symmetric under permutations 
of the qubits. This identification  leads to a family of sufficient conditions 
of inseparability based on the negativity of $2k^{\rm th}$ order  inter-group 
covariance matrices ($2k\leq N$) of  symmetric $N$-qubit systems.  
These conditions have a simple structure and detect 
{\em  entanglement in  all even partitions of the symmetric multiqubit system.} 
The  observables involved are feasible
 experimental quantities and do not demand full state 
 determination through quantum state tomography. 
\end{abstract}

\maketitle
An important problem in quantum information theory~\cite{Ple,Zyc} 
is the formulation of appropriate methods for detecting entanglement and 
then finding measures that quantify the degree of entanglement in multipartite 
systems. These two issues are difficult  to deal with in their full generality 
for examining multipartite systems and therefore, a strategy in their 
understanding is to focus on certain special symmetric states~\cite{Vol}. 
The choice of the states with specific symmetry  is based both on feasible 
experimental possibilities and on mathematical considerations~\cite{sym}. 
In this communication, we examine entanglement properties of even number of 
qubits in quantum  states obeying permutation symmetry. Symmetric multiqubit 
states form an important class due to their experimental significance~\cite{Exptl1, Exptl2}.  
Taking advantage of the elegant  mathematical structure associated with symmetric states, 
we propose a set of sufficient but not necessary conditions to detect entanglement 
via  experimentally amenable inter-particle covariance matrix. The inseparability 
conditions obtained here are generalizations of our earlier 
result~\cite{ARU} for pairwise  entanglement in symmetric multiqubit states. 
It is important to point out that in Ref.~\cite{ARU} these conditions are shown to 
exhibit a similar structure, involving the qubit cross-correlations, 
like those for the Gaussian states~\cite{Simon} and thus our approach 
reveals a structural parallelism between the continuous variable states and 
multiqubits considered here.

Symmetric multiqubit states remain invariant under any permutation of 
the qubits and are therefore restricted to a $(N+1)$ dimensional subspace 
of the entire $2^N$ dimensional Hilbert space -   allowing for a substantial 
reduction in the state space.  This  is the maximal multiplicity space of 
collective angular momentum, 
$\hat{J}_i~=~\frac{1}{2}\sum_{\mu=1}^N\, ~\sigma_{\mu\, i};\ i=x, y, z, $ 
( $\sigma_{\mu\, i}$ is the Pauli operator of $\mu$th qubit) and is 
spanned by the eigen  states 
$~\{|\frac{N}{2},M\rangle~ ;~ \,~ -\frac{N}{2}~\leq~ M~\leq ~ \frac{N}{2}~\}~$ of 
 $\hat{J}^2$ (with maximum eigenvalue $J=N/2$) and $\hat{J}_z$.  

An arbitrary  $N$ qubit system is characterized by the density matrix  
\begin{equation}
      \label{mqrho}
      \rho=\frac{1}{2^N}\, \displaystyle\sum
	    T_{\alpha_1\, \alpha_2\, \ldots  \alpha_N}\, 
	   (\sigma_{1\alpha_1}\, \sigma_{2\alpha_2}\, \ldots\sigma_{N\alpha_N})\, , 
\end{equation}
where  $\sigma_{\mu\alpha}=(I\otimes I \otimes \ldots \otimes 
\sigma_{\alpha}\otimes I \otimes \ldots),$ 
- with  $ \sigma_{\alpha}$ appearing  in the $\mu^{\rm th}$ position - denotes the
Pauli operator of the $\mu^{\rm th}$ qubit;  $\alpha_1, 
    \alpha_2,   \ldots ,  \alpha_N \ =0, x, y, z$  and 
{\scriptsize$\sigma_0~=~I=~\left( \begin{array}{rr} 1 & 0 \\ 0 & 1 \end{array} \right),  
\sigma_x~=~\left( \begin{array}{rr} 0 & 1 \\ 1 & 0 \end{array} \right), 
\sigma_y~=~\left( \begin{array}{rr} 0 & -i \\ i & 0 \end{array} \right), $
$\sigma_z~=~\left( \begin{array}{rr} 1 & 0 \\ 0 & -1 \end{array} \right);$ }
the real coefficients $T_{\alpha_1\, \alpha_2\, \ldots  \alpha_N}$ are the averages   
\begin{eqnarray} 
       \label{genT} 
       T_{\alpha_1\, \alpha_2\, \ldots  \alpha_N}&=&{\rm Tr} \left[\rho\, 
	   (\sigma_{1\alpha_1}\, \sigma_{2\alpha_2}\, \ldots 
	   \sigma_{N\alpha_N})\right]\nonumber \\ 
       &=&\left\langle \sigma_{1\alpha_1}\, \sigma_{2\alpha_2}\, 
	   \ldots \sigma_{N\alpha_N}\right\rangle ,   
\end{eqnarray} 
and $T_{0\,0\,\ldots 0}=1$  gives the normalization condition. 
Note that the total number of independent parameters in this trace-class density matrix is 
{\scriptsize$\displaystyle\sum_{r}\, \left(\begin{array}{c} N \\ r\end{array}\right)\, 3^r-1=2^{2N}-1.$}
For multiqubit states obeying exchange symmetry, the state parameters 
$T_{\alpha_1\, \alpha_2\, \ldots  \alpha_N}$ are symmetric under interchange 
of any pair of indices (corresponding to swapping of the qubits). So, the total 
number of parameters reduce to $(N+1)^2-1$. Setting $N-l$ indices  
equal to  0 and remaining $l$ indices taking values $x, y, z,$ we obtain moments 
of  $l^{\rm th}$ order $(l\leq N):$
\begin{eqnarray} 
     \label{cordef} 
      T_{i_1\, i_2\, \ldots   i_l}^{(l)}  &=& \left\langle\sigma_{1i_1}\, 
	  \sigma_{2i_2}\, \ldots     \sigma_{li_l}\right\rangle=  
      T_{i_1\,  i_2\,  \ldots \, i_l\, 0\, 0\ldots} \\
      &&      {\rm where} \  \ i_1,\, i_2,\, \ldots, i_l=x,\, y,\, z.\nonumber
\end{eqnarray}
It is convenient to introduce  collective multi-indices $i~ \equiv~ \{i_1~\,~ i_2\,~ 
\ldots~ i_k\}~,\ ~j~ \equiv ~ \{~j_1~\, ~j_2~\, ~\ldots~ j_k~\}$,  
so that the moments $T^{(2k)}$ of even order $2k$ (with  $k~=~1,\, 2,\, 
\ldots , [N/2]$)  may be arranged as   $3^k\times 3^k$ real symmetric 
matrices  and  moments $T^{(k)}$ of $k^{\rm th}$  order are arranged 
as  $3^k$ componental columns:
\begin{equation} 
       \label{2kmatrix} 
       T^{(2k)}_{i\, j}=T_{i_1\, i_2\, \ldots   i_k;j_1\, j_2\, 
	   \ldots j_k}^{(2k)}, \ \ {\rm and}\ 
	   T^{(k)}_{i}=T^{(k)}_{i_1\, i_2\, \ldots   i_k}.
\end{equation}      

Let us consider $k$ 
qubit operators $\hat{\bf A}^{(k)}$ and $\hat{\bf B}^{(k)}$, 
associated with two different  groups $a\ {\rm and}\ b$: 
\begin{eqnarray}
           \label{AB}
           \hat{A}^{(k)}_i&=&\sigma_{a_1\, i_1}\sigma_{a_2\, i_2}\ldots 
		   \sigma_{a_k\, i_k}, \nonumber \\ 
            \hat{B}^{(k)}_j&=&\sigma_{b_1\,j_1}\sigma_{b_2\, j_2}\ldots \sigma_{b_k\,j_k}. 
\end{eqnarray}
Arranging $\hat{\bf A}^{(k)}$ and $\hat{\bf B}^{(k)}$ as a column $\hat{\xi}^{(k)}$ 
of $2\cdot 3^k$ operators, (correspondingly, 
$\hat{\xi}^{(k)^\dag}~=~(\hat{\bf A}^{(k)\dag},\ \hat{\bf B}^{(k)\dag}),$ as a row of operators),  
we define the $2k^{\rm th}$ order  variance matrix, as in Ref.~\cite{ARU}
\begin{equation}
	\label{var}
	{\cal V}^{(2k)}=\frac{1}{2}\, \left[\left\langle\hat{\Delta}\xi^{(k)}\, 
	\hat{\Delta}\xi^{(k)\dag}\right\rangle+ \ {\rm h.c}\right] ,
\end{equation}
where $\hat{\Delta}\xi^{(k)}=\hat{\xi}^{(k)}-\langle\hat{\xi}^{(k)}\rangle$. 
Note that by construction (\ref{var}),  ${\cal V}^{(2k)}$ is  
 $(2\cdot 3^k\times 2\cdot 3^k)$ dimensional real symmetric positive semi definite matrix.
The elements of the variance matrix are  
\begin{equation}
          {\cal V}^{(2k)}_{ij}=\frac{1}{2}\, \left(\langle\{\hat{\xi}^{(k)}_i,\, \hat{\xi}^{(k)}_j\}\rangle 
          -\{\langle\hat{\xi}^{(k)}_i\rangle ,\,\langle \hat{\xi}^{(k)}_j\rangle\}\right)
\end{equation} 
 ( where $\{\hat{\xi}^{(k)}_i,\, \hat{\xi}^{(k)}_j\}~=~\hat{\xi}^{(k)}_i\, \hat{\xi}^{(k)}_j~+~\hat{\xi}^{(k)}_j\, 
\hat{\xi}^{(k)}_i $) and ${\cal V}^{(2k)}$ is cast in a $(3^k\times 3^k)$ block form, 
\begin{equation}
	\label{variance}
          {\cal V}^{(2k)}=\left(\begin{array}{cc} {\cal A}^{(2k)} & {\cal C}^{(2k)} \\ 
          {\cal C}^{(2k)T} &   {\cal B}^{(2k)} 
		  \end{array}\right). 
\end{equation}
Clearly, the off-diagonal block ${\cal C}^{(2k)}$ corresponds to  
$2k^{th}$ order covariances  among the inter-group of multiqubits
\begin{eqnarray}
	\label{cmatrix}  
          {\cal C}_{ij}^{(2k)}&=&\langle \hat{A}^{(k)}_i\, \hat{B}^{(k)}_j\rangle - 
          \langle\hat{A}^{(k)}_i\rangle\,\langle\hat{B}^{(k)}_j\rangle  \nonumber \\
          &=& T^{(2k)}_{i\, j}-T^{(k)}_i\, T^{(k)}_j.  
\end{eqnarray}
In the second line of (\ref{cmatrix}), we have  used (\ref{cordef}) and  (\ref{2kmatrix}).
 The diagonal blocks ${\cal A}^{(2k)}$  and ${\cal B}^{(2k)}$ are identical for 
 a symmetric  intra-group multiqubit system: {\scriptsize ${\cal A}^{(2k)}_{ij}
={\cal B}^{(2k)}_{ij}~=~\langle \hat{A}^{(k)}_i\, ~\hat{A}^{(k)}_j\rangle ~
~ -~ \langle\hat{A}^{(k)}_i~\rangle\, ~\langle\hat{A}^{(k)}_j\rangle~$} because 
the intra-group averages are the same viz., {\scriptsize$\langle \hat{A}^{(k)}_i\, ~ \hat{A}^{(k)}_j\rangle ~
=~\langle \hat{B}^{(k)}_i\, ~ \hat{B}^{(k)}_j\rangle$} and {\scriptsize$\langle \hat{A}^{(k)}_i\rangle 
~=~\langle \hat{B}^{(k)}_i\rangle~.$}
Under {\em identical}
local unitary transformations $U\otimes U\otimes \ldots \otimes U$ on the 
qubits - which preserve the symmetric space structure -  the blocks  of the 
variance matrix change as
 \begin{eqnarray} 
	 {\cal A}^{(2k)}\rightarrow {\cal A}^{(2k)'}&=&{\cal R}\, {\cal A}^{(2k)}{\cal R}^T,\nonumber \\ 
	 {\cal C}^{(2k)}\rightarrow {\cal C}^{(2k)'}&=&{\cal R}\, {\cal C}^{(2k)}{\cal R}^T,
 \end{eqnarray} 
  with  {\scriptsize${\cal R}~=~\begin{array}{c} \underbrace{R\otimes R\otimes \ldots \otimes R} 
 \\ k\ {\rm times}\end{array}$},  a $3^k\times 3^k$ 
 real orthogonal matrix, comprised of direct products (containing $k$ factors) 
 of 3 dimensional rotations $R\in {\rm SO(3)}$ - corresponding uniquely to  
 $2\times 2$ unitary matrices $U\in SU(2)$.            

We now focus on the question: {\em How would multiqubit entanglement 
manifest itself under different partitioning of a symmetric system? }
Our identification here is  that  the inter-group covariance matrix  ${\cal C}^{(2k)}$ 
holds a key to symmetric multiqubit entanglement,  coming from various {\em even} partitioning of the system.  
It is worth mentioning at this juncture the important difference between the recent paper of 
Korbicz et. al.~\cite{Korbicz}  from our present work. These authors have proposed 
necessary and sufficient  conditions for entanglement, reflected through  
 two and three qubit partitions  of a symmetric multiqubit system. 
Strikingly, the two qubit result is shown~\cite{ARU} to be captured by the off-diagonal 
block of the variance matrix. An important open problem, concerning the inseparability 
features hidden in {\em all the even qubit reduced systems} of a symmetric $N$-qubit state, 
 is what we are addressing here, by generalizing our approach  
outlined in Ref.~\cite{ARU}.  

First of all, we note that  positivity of the variance matrix 
${\cal V}^{(2k)}$ demands that the diagonal blocks 
${\cal A}^{(2k)}$  be positive semidefinite. However,  there are no 
constraints of positivity on the off-diagonal block ${\cal C}^{(2k)}$ 
as such, though separable symmetric states carry a distinguishing feature:   
 
{\bf Theorem}:  For every separable symmetric multiqubit state, inter-group covariance 
matrices ${\cal C}^{(2k)}$ of various order $2k\leq N$ are necessarily positive semidefinite. 
 
{\bf Proof:} Consider a separable symmetric state of $2k$ qubits,  which is 
decomposable as a convex sum of direct products of $k$ qubit density matrices $\rho^{(k)}_w$:
 \begin{equation}
	 \label{ksep}
	 \rho^{(2k)}_{\rm sep}= \displaystyle\sum_w p_w \, \rho^{(k)}_w\otimes \rho^{(k)}_w; \ \ 
	 \displaystyle\sum_{w}p_w=1; \ \ 0\leq p_w\leq 1.
 \end{equation}
In this state  the inter-qubit averages $\langle \hat{A}^{(k)}_i\, \hat{B}^{(k)}_j\rangle$  
are also separable:
\begin{eqnarray}
	\langle \hat{A}^{(k)}_i\, \hat{B}^{(k)}_j\rangle_{\rm sep} &=&
	\displaystyle\sum_{w} p_w\, \langle\hat{A}^{(k)}_i\rangle_w \,				                 
	\langle\hat{B}^{(k)}_j\rangle_w \nonumber \\ 
	&=& \displaystyle\sum_{w} p_w\, \langle\hat{A}^{(k)}_i\rangle_w \, 					                 
	\langle\hat{A}^{(k)}_j\rangle_w \nonumber \\ 
	&=&\displaystyle\sum_{w} p_w\, T^{(k)}_{i}(w)\, T^{(k)}_{j}(w), 
\end{eqnarray}  
where we have denoted ${\rm Tr}(\rho^{(k)}_w~\, ~\hat{A}^{(k)}_i)~=~\langle\hat{A}^{(k)}_i~\rangle_w~$ and 
used the fact $\langle\hat{A}^{(k)}_i~\rangle_w~=~\langle\hat{B}^{(k)}_i~\rangle_w$ 
and  the notation  $\langle\hat{A}^{(k)}_i~\rangle_w~=~T^{(k)}_i(w)$. 
It is also clear that
\begin{equation} 
	\langle \hat{A}^{(k)}_i\rangle_{\rm sep}=\langle \hat{B}^{(k)}_i\rangle_{\rm sep}=
	\displaystyle\sum_{w} p_w\, T^{(k)}_{i}(w).
\end{equation}

The real quadratic form
{\scriptsize$Q^{(2k)}=X^T\, {\cal C}^{(2k)}\, X=\sum_{i,j}{\cal C}^{(2k)}_{ij}\, X_i\,  X_j,$}
 with an arbitrary  real $3^k$ componental column $X$, when evaluated in 
 the separable state (\ref{ksep}) gives  
 \begin{equation}
	 Q^{(2k)}_{\rm sep}=\displaystyle\sum_{w} p_w\, \left(T^{(k)}_{i}(w) \, X_i\right)^2-
	\left( \displaystyle\sum_{w} p_w\, \left(T^{(k)}_{i}(w) \, X_i \right)\right)^2
 \end{equation} 	
which is necessarily a positive semidefinite quantity,  implying in turn  that                 
 ${\cal C}^{(2k)} \geq 0;\ k=1,2,\ldots , [N/2],$ in a  separable symmetric 
 multiqubit  state (\ref{ksep}). This proves our theorem. $\Box$
 
The above theorem leads to sufficient conditions for  entanglement associated with
 even number of qubits in a symmetric state: 
{\em If the inter-group covariance matrix ${\cal C}^{(2k)}$ is negative, then the 
symmetric multiqubit state exhibits $2k$-qubit entanglement} for $k=1,2,\ldots $ with $2k~\leq ~ N$. 
This   leads to a hierarchy of inseparability conditions, 
which test entanglement in even partitioning.  
For $k=1$, the condition  $C^{(2)}<0$   has been shown in Ref.~\cite{ARU} 
to be a direct consequence of Peres-Horodecki partial transpose criterion
~\cite{Peres} on  two-qubit partitions of a  symmetric multiqubit state. 
Thus, negativity  of the covariance matrix $C^{(2)}$ serves as  both  
necessary and sufficient for pairwise  entanglement in the symmetric $N$ qubit system. 
Any test which confirms the negativity of the real symmetric $3^k\times 3^k$ covariance matrix 
${\cal C}^{(2k)}$ is sufficient to assert the inseparability of the symmetric multiqubit state.  
In order to establish the negativity of  ${\cal C}^{(2k)},$ the Sylvester 
criterion~\cite{Syl} may be used:  {\em Negative value assumed by any of the 
principal minors of a hermitian matrix implies that the matrix is not positive semidefinite}. 
Thus a series of sufficient conditions  for entanglement of $2k$ qubits could be 
extracted from negative principal minors  (of various orders)~\cite{eigenvalues} 
of the corresponding covariance matrix  ${\cal C}^{(2k)}$. This brings out inseparability 
constraints involving a few correlation observables, making our criterion experimentally 
amenable. It may be noted that  a series of inseparability conditions, resulting 
from negative principal minors of various orders, demonstrate~\cite{SV}
negativity of the (infinite dimensional) partial transpose of a bipartite continuous 
variable density matrix, which is the Peres-Horodecki criterion~\cite{Peres} for infinite 
dimensional states. 
  
We now test our  inseparability conditions ${\cal C}^{(2k)}<0$ by considering some 
well known examples of symmetric $N$-qubit  states such as,  GHZ and  W type states, 
which have attracted experimental focus~\cite{Exptl1}.  

For an even~\cite{oddN}   $N$ qubit GHZ state~\cite{Dur}: 
{\scriptsize \begin{equation} 
    \label{ghz}
	  \vert {\rm GHZ}_N\rangle = 
	  \frac{1}{\sqrt{2}} \, ( \vert 0_N\rangle 
	 + \vert 1_N\rangle )=
	 \frac{1}{\sqrt{2}} \, ( \vert 0\, 0\, \ldots 0\rangle 
	  + \vert 1\, 1\, \ldots 1\rangle )  
\end{equation}}
we find that $C^{(N)}$ has one negative eigenvalue, ~$~\lambda^{(-)}~=~-2^{(\frac{N}{2}-1)}~$. 
The lower order covariances ${\cal C}^{(2k)}$, for $k<N/2$,  are all positive semidefinite. 
This is obvious because  GHZ state is separable with the disposal of qubits.  
Thus, our ${\cal C}$-matrix criterion is in concordance with  the known result that the  
{\em {\rm GHZ} state is  $N$-party entangled  and  is fragile under 
disposal of particles}~\cite{Dur}.  

From  experimental point of view, it may be noted that the lowest order 
(see ~\cite{eigenvalues}) principal minor, which records negativity of 
$C^{(N)}$ is the diagonal element, with the index $i=\{xxx\ldots xy\}$: 
{\scriptsize\begin{equation}
   \label{cnghz}
         C^{(N)}_{ii}=T^{(N)}_{ii}-\left(T^{(\frac{N}{2})}_i\right)^2 
         =\left\{ \begin{array} {l}
         -1, \ \ {\rm if \ \ N/2=even \ integer,}\\
         -2, \ \ {\rm if \ \ N/2=odd \ integer.}\\  
\end{array}\right.
\end{equation}}
 More specifically, even-$N$ qubit entanglement in GHZ states is 
 revealed~\cite{ghzfn} by the measurement of the $N$-qubit observable 
{\scriptsize$\langle\sigma_{1\,x}\sigma_{2\,x}\ldots\sigma_{\frac{N}{2}\,y}
\sigma_{\frac{N}{2}+1\,x}\sigma_{\frac{N}{2}+2\,x}\ldots\sigma_{N \, y}\rangle$}  
(where the qubit indices may be conveniently interchanged). 
 
Next,  consider  $N$  qubit W-state~\cite{Dur}:
{\scriptsize\begin{equation} 
      \label{W}
	  \vert {\rm W}_N\rangle = \frac{1}{\sqrt{N}} \, 
	  ( \vert 1\, 0\, 0\, \ldots 0\rangle + \vert 0\, 1\, 0\, \ldots 0\rangle+\ldots).  
\end{equation}}
Here, the covariance matrices ${\cal C}^{(2k)}$, of all orders 
$~k=~1,2,\ldots, \ [N/2], ~$ are negative ( with only one negative eigenvalue,  
$\lambda ^{(-)}=-\frac{2k(k-1)}{N^2}$). Therefore,  {\em W-state of $N$-qubits 
is confirmed to exhibit $2k$ qubit entanglement for all values of $k$  
(with, of course,   $2k\leq N$}). Here again, the $2k$ qubit entanglement 
is  seen explicitly through  the measurement of one of the diagonal elements 
of the covariance matrix ${\cal C}^{(2k)}_{ii}$, with $i=\{zzz\ldots z\}$, for 
which ${\cal C}^{(2k)}_{ii}=T_{ii}^{(2k)}=-1$. It is therefore sufficient  
to check that  $\langle \sigma_{1\,z}\sigma_{2\,z}
\ldots \sigma_{2k\,z}\rangle$ is {\em negative}. Thus the  W state 
has $2k$ qubit entanglement in all the even partitions $2k=2,4,6,\ldots $  of the state. 
Our results confirm that  {\em the $W$ state is robust under disposal of qubits}~\cite{Dur}. 

We now investigate  the implications of our inseparability conditions 
${\cal C}^{(2k)}<0$ for mixed states:  To this end,  suppose that experimentally produced 
W and GHZ states have  noise-like admixture of  incoherently superposed 
symmetric states:
{\scriptsize\begin{equation}
       \label{mixedstate}
        \rho_{\rm noisy}^{(N)}=\frac{(1-x)}{N+1}\, P_N+\,x\, 
		\left\vert\psi\right\rangle\left\langle\psi\right\vert
		, \ \ 0\leq x \leq 1,
\end{equation}}
where  {\scriptsize$P_N=\displaystyle\sum_{M=-\frac{N}{2}}^{\frac{N}{2}}\, 
\left\vert\frac{N}{2} \, M\right\rangle\, \left\langle\frac{N}{2}\, 
M \right\vert$} denotes  the projection operator  onto the symmetric 
subspace ${\rm Sym}~\,(C^2\otimes ~C^2\otimes ~\ldots ~\otimes ~C^2)~$ 
of $N$ qubits ($P_N$ is an identity matrix in the symmetric subspace of 
qubits and hence {\scriptsize$P_N/(N+1)$} corresponds to a  maximally disordered  
separable symmetric  state),   and $\vert\psi\rangle$ is either 
a $N$-qubit  $|{\rm GHZ}\rangle_N$  or $|{\rm W}\rangle_N$ state. For the 
least eigenvalue of the covariance matrix $C^{(N)}$ to be negative, 
the mixing parameter $x$ has to be greater than a certain threshold value. 
We find the following range of $x$ for which inseparability is indicated 
via negativity of $C^{(N)}$ for N=2, 4 and 6 qubits: 
{\scriptsize \begin{eqnarray}  
	 \label{mixingparameterghz}
	{\rm GHZ-noisy\ state:}\ \     0.25< x \leq  1, &{\rm for}\  N=2. \nonumber\\ 
	 0.0625< x\leq 1,&{\rm for}\  N=4.\nonumber \\
	0.014 < x\leq 1,  &{\rm for}\  N=6. \\
	\label{mixingparameterw}
	{\rm W-noisy\ state:}\ \   0.25<x\leq 1,  &{\rm for}\  N=2. \nonumber \\ 
	 0.0899<x \leq 1, & {\rm for}\  N=4. \nonumber \\ 
	 0.042 < x \leq 1,  &{\rm for}\  N=6. 
\end{eqnarray}}
We observe that the $x$-range for  $N$-qubit 
entanglement is smaller for the noisy W state (see (\ref{mixingparameterw})), 
 than that  (\ref{mixingparameterghz}) for the noisy GHZ state. But eventually for 
large $N$, both  the noisy states remain entangled  for all values of $x$.
A more general trend  (but a restricted domain for $x$) is found by 
examining the lowest order principal minor:  the noisy GHZ state  is $N$ (even) 
 qubit entangled, when {\scriptsize$\frac{1}{N^2}<x \leq  1$},  ( verified by demanding 
 that the diagonal element {\scriptsize$T^{(N)}_{ii}=\frac{(1-x)}{(N^2-1)}-x<0;$} 
the index $i=\{x\, x\, \ldots \, \, y\}$). For mixed noisy state of W, the 
inseparability range - for $N$ qubit entanglement - is identified to 
be  {\scriptsize$\frac{1}{(N+2)}<x\leq 1,$} resulting  from the negative diagonal element
{\scriptsize$T^{(N)}_{ii}~=~\frac{(1-x)}{(N+1)}-x$}, with $i=\{z\, z\, \ldots \, z\}.$

Entanglement in  various even partitions of the W-noisy state (\ref{mixedstate}) 
are examined by using  the $n$ qubit reduced W noisy state,
{\scriptsize\begin{eqnarray}
	\label{WN-n}
	\rho^{(N-n)}_{\rm noisy}(W)\hskip .06 in=\hskip .15 in \frac{(1-x)}{(N-n+1)}\, P_{N-n}\hskip 1.02 in \nonumber\\ 
	 \ \ +\,x\,\left[\frac{(N-n)}{N}\, \vert{\rm W}_{N-n}\rangle\,  \langle{\rm W_{N-n}}\vert +\frac{n}{N}\, \vert 	
	0_{N-n} \rangle\,\langle 0_{N-n}\vert\right].\hskip 0.07in
\end{eqnarray}}
The covariance matrices of  all  even partitions of the W noisy state are 
found to be negative, in  a specific inseparability range of  the mixing parameter $x$.  
For example we find that a W noisy state is two-qubit entangled when 
{\scriptsize$\frac{N^2}{N^2+12}<x\leq 1$}. Note that this inseparability range for two-qubit entanglement 
 is much restricted than the one realized for entanglement in the largest even partition of the state 
 (see (\ref{mixingparameterw})).  As $N$ increases $x\rightarrow 1$ indicating 
 that in the large $N$ limit the two qubit partition of a noisy W state  is separable 
 throughout the range $0\leq x<1.$ The  $n$ qubit reduced GHZ noisy state is a convex 
 sum of three separable states {\scriptsize$P_{N-n}/(N-n+1)$, $\vert 0_{N-n} \rangle\, \langle 0_{N-n}\vert$} and 
{\scriptsize$\vert 1_{N-n} \rangle\, \langle 1_{N-n}\vert$} and is thus  a separable state. 

In conclusion, we have here generalized our formalism of the symmetric 
two-qubit inseparability condition, expressed in terms of inter-qubit 
covariance matrix~\cite{ARU}, to all even qubit partitions of symmetric 
$N$-qubit systems. This takes the form of a hierarchy of inseparability 
conditions on the inter-group covariance matrices of even order:  
${\cal C}^{(2k)}<0$ , $k=1,\,2,\,\ldots$ with $2k\leq N$. Only for $k=1$ 
(i.e., for two qubit partitions) the inseparability condition 
is both necessary and sufficient, and for all other values of $k$, these 
conditions are only sufficient. We have illustrated their use for both pure 
and mixed states involving GHZ and W type states.
The symmetric multiqubit system considered here 
facilitates a  richer analysis in terms of SO(3) irreducible tensors~\cite{ARU2}. 
The irreducible tensor approach leads to a family of criteria~\cite{ARU2} for 
entanglement based on covariance matrices involving   collective angular momentum 
variables and is suitable to test inseparability  in macroscopic atomic 
ensembles~\cite{Exptl2}. Our approach suggests  further generalization  
to $d$-level symmetric multiparticle systems also.


\begin{thebibliography}{1}
\bibitem{eigenvalues} While one can directly check negativity of the finite 
                      dimensional matrix ${\cal C}^{(2k)}$ by evaluating negative 
					  eigenvalues of the matrix (if they exist), using standard linear 
					  algebra packages, experimental  verification of a negative 
					  eigenvalue would involve larger number of measurements. 
					  Easily implementable experimental test of entanglement results from 
					  lowest order negative principal minor.
\bibitem{oddN}     Though our criterion is not applicable to detect 
                   $N$-qubit entanglement for odd-$N$, we note that a $(N-1)$ qubit 
				   reduced system, with one qubit removed,  is verifiable by our methods. 
				   A genuine $N$-party  entangled state (with odd-$N$) is separable with 
				   the disposal of a qubit and this could be tested through positive semidefinite 
                   covariance matrices ${\cal C}^{(2k)}$ for all $2k=N-1, N-3,\ldots , 2$. However, 
				   verification of positivity of  ${\cal C}^{(2k)}$ does not ensure that the 
				   state is separable, as this criterion is only sufficient.
\bibitem{ghzfn}  The diagonal element $C_{ii}^{(N)}=T_{ii}^{(2k)}-(T_i^{(k)})^2<0$ whenever 
                 $T_{ii}^{(2k)}$ itself is negative. Negative value assumed by  any 
				 diagonal element $T_{ii}^{(2k)}$ is therefore sufficient to check 
				 negativity of ${\cal C}^{(2k)}$. For example, in an even $N$ qubit 
				 GHZ state,  $~T_{ii}^{(N)}~=~\langle\sigma_{1\,x}\sigma_{2\,x}\ldots
                 \sigma_{\frac{N}{2}\,y}\sigma_{\frac{N}{2}+1\,x}\sigma_{\frac{N}{2}+2\,x} 
				 \ldots\sigma_{N \, y}\rangle~=~-~1~$, (where the multi-index 
				 $i~=~ \{ x\, x\, \ldots \, y\})$, while the $N/2$-qubit correlation 
				 observable $\langle\sigma_{1\,x}\sigma_{2\,x}\ldots\sigma_{\frac{N}{2}\,y}\rangle$ 
				 takes the value 0, 1  depending on whether $N/2=$ even or odd respectively. 
				 Measurement of  $N$ qubit correlation $T_{ii}^{(N)}=\langle\sigma_{1\,x}
                 \sigma_{2\,x}\ldots\sigma_{\frac{N}{2}\,y}\sigma_{\frac{N}{2}+1\,x}
                 \sigma_{\frac{N}{2}+2\,x} \ldots\sigma_{N \, y}\rangle$  is  therefore  
				 sufficient  to verify multiqubit entanglement.
\end{thebibliography}
 \end{document}